\documentclass[preprint,aps,showpacs,nofootinbib,tightenlines,showkeys,
    byrevtex]{revtex4}
\usepackage{graphicx}%
\usepackage{dcolumn}
\usepackage{amsmath}
\usepackage{epsfig}
\usepackage{latexsym}
\usepackage{amssymb}
\usepackage{mathrsfs}

\newcommand{\dis}{\displaystyle}
\newcommand{\ii}{\mathrm{i}}
\newcommand{\dd}{\mathrm{d}}
\newcommand{\ia}{{\"{\i}}}

\begin{document}
\bibliographystyle{plain} 

\preprint{Preprints nucl-th/0207034, LBL-51037, TUM-T39-02-12}

\title{Low Energy Expansion in the Three Body System to All Orders and the
  Triton Channel}

\author{Paulo F.~Bedaque}\email{pfbedaque@lbl.gov} \author{Gautam
  Rupak}\email{grupak@lbl.gov} \affiliation{Nuclear Science Division, Lawrence
  Berkeley National Laboratory, Berkeley, CA 94720, USA}

\author{Harald W.~Grie\ss hammer}\email{hgrie@physik.tu-muenchen.de}
\affiliation{Institut f\"ur Theoretische Physik (T39), Physik Department der
  Technischen Universit\"at M\"unchen,
  D-85747 Garching, Germany\\
  \textnormal{and}\\
  ECT*, Villa Tambosi, I-38050 Villazzano (Trento), Italy
  }

\author{H.-W.~Hammer}\email{hammer@mps.ohio-state.edu}
\affiliation{Department of Physics, The Ohio State University, Columbus, OH
  43210, USA}

\date{15th July 2002, revised 24th October 2002}

\begin{abstract}
  We extend and systematise the power counting for the three-body system, in
  the context of the ``pion-less'' Effective Field Theory approach, to all
  orders in the low-energy expansion.  We show that a sub-leading part of the
  three-body force appears at the third order and delineate how the expansion
  proceeds at higher orders. After discussing the renormalisation issues in a
  simple bosonic model, we compute the phase shifts for neutron-deuteron
  scattering in the doublet $\mathrm{S}$ wave (triton) channel and compare our
  results with phase shift analysis and potential model calculations.
\end{abstract}



\maketitle

%%%%%%%%%%%%%%%%%%%%%%%%%%%%%%%%%%%%%%%%%%%%%%%%%%%%%%%%%%%%%
\begin{section}{Introduction}
\label{introduction}
%%%%%%%%%%%%%%%%%%%%%%%%%%%%%%%%%%%%%%%%%%%%%%%%%%%%%%%%%%%%
The disparity between the energy scales of typical QCD phenomena and the scale
for nuclear binding makes nuclear systems an ideal playing ground for
Effective Field Theory (EFT) methods. The first motivation behind the EFT
program in nuclear physics is to obtain a description of nuclear phenomena
firmly based on QCD.  Recently, considerable effort has been put into the
development of a phenomenologically successful EFT for nuclear
physics~\cite{bira_review, seattle_review,bedaque_bira_review,rho_review}.  It
soon became apparent that there is one feature which distinguishes the nuclear
EFT from most of the many other applications of EFT methods: The low-energy
expansion is non-perturbative in the sense that an infinite number of diagrams
contribute at each order. This follows from the very existence of shallow real
and virtual bound states whose binding energies are so small that they lie
within the range of validity of the EFT even though they cannot be described
by perturbing the free theory. In fact, there are at least in the two and
three-nucleon sector states so loosely bound that their scales do not seem to
be connected even to the soft QCD scales $m_\pi$ or $f_\pi$. Instead, they
appear to require an additional ``accidental'' fine-tuning whose microscopic
origin is not understood. However, this phenomenon offers the opportunity for
a more radical use of EFT in the few nucleon case. In this approach, all
particles but the nucleon themselves are considered high energy degrees of
freedom and are consequently ``integrated out''. The resulting EFT is
considerably simpler than potential models or the ``pionful'' version of
nuclear EFT (in which pions are kept as explicit degrees of freedom), but its
range of validity is reduced to typical momenta below the pion mass.  Albeit
this might seem a severe restriction, there are many processes situated in
this range which are both interesting in their own right and important for
astrophysical applications. Also, we believe that an understanding of this
simpler EFT is a necessary condition for understanding more general
applications of EFT to nuclear physics and to other systems which exhibit
shallow bound states.

The ``pionless'' EFT discussed here has its historical roots in the effective
range expansion~\cite{Bethe} and in the model-independent approach to
three-body physics~\cite{efimovI}.  It considerably extends those approaches
by making higher order calculations systematic, by including external
electroweak currents, and by providing rigorous error estimates.  A number of
technical issues like gauge invariance and the inclusion of relativistic
corrections are also much simpler in this approach.  To gain perspective, it
is useful at this point to briefly review what has been accomplished towards
the ``pionless'' EFT program.  In the two-nucleon sector, the modification of
the power counting needed to deal with the fine tuning alluded to above and
the resulting non-perturbative nature of the low-energy expansion is well
understood.  It has been phrased in elegant terms using dimensional
regularisation which allows for analytical calculations. A large number of
processes were analysed, some of them to very high order (see
Refs.~\cite{bira_review,seattle_review,bedaque_bira_review,rho_review} and
references therein).  Typically, most of the low-energy constants required are
obtained from the plethora of available nucleon-nucleon scattering data.  At
some but usually high order, new two-body operators appear which describe
effects due to ``exchange currents'', ``quark effects'', etc. They contribute
typically in the range of a few percent to the final result.  The coefficients
of these operators are not constrained by nucleon-nucleon scattering data, but
have to be fit to experiments in which the interaction of external currents
with the two-nucleon system is probed. The low-energy expansion converges very
nicely, the computations are not excessively complicated, and the only
obstacle in obtaining arbitrary accuracy is the question whether there are
enough data available to fix the low-energy constants.

In contradistinction, we are in the three-body sector still at a more
primitive stage. For most channels of the three-nucleon system (without
external currents), the approach used in the two-nucleon sector can be
extended in a straightforward way.  Analytical calculations, however, cannot
be performed any more, as the basic Faddeev equation is no longer soluble in
closed form. All these channels are repulsive, so that consequently no bound
states exist, and the only observables are the nucleon-deuteron phase shifts
which have been computed to third
order~\cite{2stooges_quartet,3stooges_quartet,pbhg,chickenpaper}.  The only
operators not determined by nucleon-nucleon data contributing to
nucleon-deuteron scattering are three-body forces which are highly suppressed
in these repulsive channels.  In contradistinction, the $^3$He-$^3$H channel
exhibits the effect of the two-nucleon fine tuning much more dramatically.
This leads to a very unusual situation. On a diagram-by-diagram basis, the
graphs contributing at leading order (LO) to nucleon-deuteron scattering are
finite. This seems to agree with the expectation, based on na{\ia}ve
dimensional analysis, that the three-body force, introduced to absorb any
divergence, does not appear at leading order. However, it turns out that the
ultraviolent behaviour of the equation describing the three-body system is
very different from the behaviour of its perturbative expansion. In
particular, a non-derivative three-body force is required to absorb cutoff
dependences already at leading order. This immediately voids the use of the
effective range expansion. The precise value of the three-body force is not
determined by nucleon-nucleon data and has to be determined by some additional
three-nucleon datum. These unusual properties of the Faddeev equation and the
existence of a new free parameter in the three-body sector was recognised
early on~\cite{danilov}, and the renormalisation theory interpretation was
given in~\cite{3stooges_boson}.  These complications hindered the extension of
a consistent and economical power counting to the three-body sector, and
currently only leading and next-to-leading order (NLO) calculations of the
doublet $\mathrm{S}$-wave neutron deuteron phase shift and of the
corresponding triton binding energy are
available~\cite{3stooges_doublet,doubletNLO}.  To attain the precision
necessary to be of phenomenological interest, one has to go at least one order
beyond that.

This paper is devoted to re-formulating the renormalisation issues in the
$^3$He-$^3$H sector in order to extend the power counting to all orders. In
the following section, we do so first in a bosonic model to avoid unrelated
complications coming from the spin-isospin structure. In
Sect.~\ref{sec:triton}, we consider the three-nucleon system, and compute then
the neutron-deuteron phase shift in this channel at next-to-next-to-next
leading order (NNLO) in Sect.~\ref{sec:numerics}. There, we also compare the
results with the available phase shift analysis~\cite{doublet_PSA}, and with
potential model calculations~\cite{doublet_pots}.  This also allows us to
discuss the convergence of the low-energy expansion in order to obtain a
credible error estimate. We close by conclusions, outlook and an appendix.
              %An Appendix treats issues suitable for Appendices.

%%%%%%%%%%%%%%%%%%%%%%
\section{Bosonic interlude: Power counting to all orders}

Let us first consider the renormalisation of higher order corrections in the
simpler setting of the model of three spinless bosons of mass $M$ whose two
body system exhibits a shallow real bound state. This avoids the technical
complications with spin and isospin degrees of freedom. The Lagrangean is thus
\begin{equation}\label{boson_lag}
\begin{split}
  {\mathscr L}=\psi^\dagger\left(\ii\partial_0+\frac{\nabla^2}{2M}\right)\psi
  &+\frac{C_0}{4}|\psi|^4 +
  \frac{C_2}{4}\left((\psi^\dagger(\overrightarrow{\nabla}
    -\overleftarrow{\nabla})^2\psi^\dagger)\psi^2
    +\text{H.c.}\right)\hfill\\
  &+\hfill\frac{D_0}{36}|\psi|^6
  +\frac{D_2}{36}\left((\psi^\dagger(\overrightarrow{\nabla}
    -\overleftarrow{\nabla})^2\psi^\dagger)\psi^\dagger\psi^3
    +\text{H.c.}\right) +\dots\;\;,
\end{split}
\end{equation}
where the terms not explictly shown are further suppressed. A more convenient
form of writing the same theory is obtained by introducing two dummy fields
$d$ and $t$ with the quantum numbers of two and three particles (referred to
in this section as the ``deuteron'' and ``triton'')
\begin{equation}\label{boson_deuteron_lag}
\begin{split}
  {\mathscr L}=\psi^\dagger\left(\ii\partial_0+\frac{\nabla^2}{2M}\right)\psi
  &+d^\dagger\left(-\ii\partial_0-\frac{\nabla^2}{4M}+\Delta\right)d
  +t^\dagger\left(\ii\partial_0+\frac{\nabla^2}{6M}+\frac{\gamma^2}{M}+
    \Omega\right)t\\
  &-\frac{g}{\sqrt{2}}\left( d^\dagger\psi^2 +\text{H.c.} \right) -\omega
  \left( t^\dagger d\psi +\text{H.c.} \right) +\dots\;\;.
\end{split}
\end{equation} The equivalence between the two Lagrangeans
(\ref{boson_lag}) and (\ref{boson_deuteron_lag}) is established by performing
the Gau{\ss}ian integral over the auxiliary fields $d$ and $t$ and
disregarding terms with more fields and/or derivatives, that are suppressed
further~\cite{pbhg}. However, there is a subtlety. In order to eliminate time
derivatives and express the result of this integration in a manifestly boost
invariant way as in (\ref{boson_lag}), one has to perform a field redefinition
\begin{equation}
\psi\rightarrow\psi + \frac{g^2}{2 \Delta^2}
|\psi|^2 \psi
+\left( \frac{7 g^4}{8 \Delta^4}-\frac{3 g^2
\omega^2}{4 \Delta^2\Omega^2}\right)
 |\psi|^4 \psi\;\;.
\end{equation} The relation between the constants in
the two Lagrangeans is then given by
\begin{equation}\label{rel_constants}
\begin{split}
  C_0 &= -\frac{2 g^2}{\Delta}\;\;,\\
  C_2 &= \frac{1}{4M}\frac{g^2}{\Delta^2}\;\;,\\
  D_0 &=-\frac{18 g^2 \omega^2}{\Omega \Delta^2}\;\;,\\
  D_2 &= \frac{21g^2}{4M\Delta^4}-\frac{9g^2 \omega^2}{2M
      \Omega^2\Delta^2} \;\;.
\end{split}
\end{equation}
We can identify three scales in the problem: $Q$ is a typical external
momentum, $R$ is the short distance scale in position space beyond which the
EFT breaks down, and $\gamma=\sqrt{MB_2}$ the anomalously small binding
momentum of the shallow two boson bound state of binding energy $B_2$. This
scale is, up to effective range corrections, equal to the inverse scattering
length in the two-body channel.  We thus perform an expansion in powers of two
small dimensionless numbers, $Q R\ll 1$ and $\gamma R\ll 1$, while keeping
terms with arbitrary powers of $Q/\gamma\sim 1$. The corresponding power
counting has been extensively discussed elsewhere and will not be repeated
here, see e.g.~\cite{bira_review,seattle_review,bedaque_bira_review} and
references therein.  Its details depend on the particular regularisation and
renormalisation scheme used.  In this work, we use a sharp momentum cutoff to
regulate divergent integrals\footnote{At present, it is not known how to
  extend dimensional regularisation, which is the scheme used in two-body
  calculations, to the three-body sector.}.  The result of these
considerations can be summed by stating which diagrams contribute at each
order in the low-energy expansion.

The bare deuteron propagator is given by the constant term $\ii/\Delta$. At
leading order (LO), the $d$ field propagator is given by the sum of \emph{all}
graphs made up from the interaction proportional to $g$ in
(\ref{boson_deuteron_lag}) with an arbitrary number of loops, see
Fig.~\ref{2body}.
%%%%%%%%%%%%%%%%%%%%%%%%%%%%%%%%%%%%%%%%%%%%%%%%%%%%%%%%%%
\begin{figure}[!htb]
\begin{center}
  \includegraphics[width=0.8\linewidth,angle=0,clip=true]{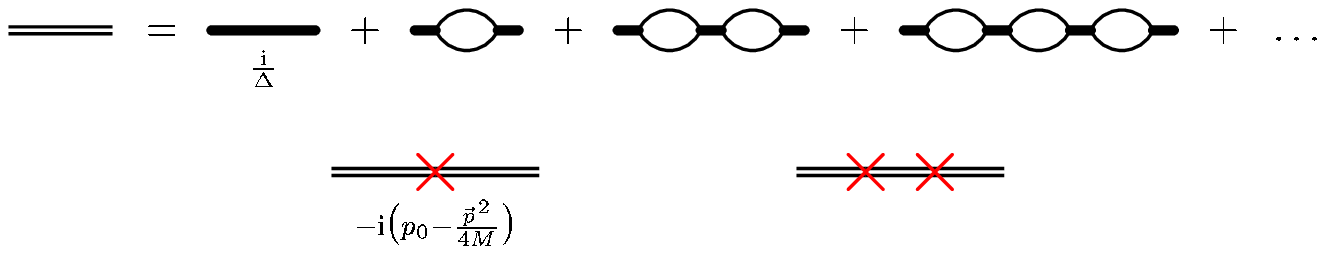}
\end{center}
\vspace*{-0pt}
\caption{Top: Re-summation of the bare deuteron propagator (thick solid line)
  into the dressed deuteron propagator (double line) at LO by dressing with
  two-nucleon bubbles. Bottom: The NLO and NNLO corrections to the deuteron
  propagator. The cross denotes an insertion of the deuteron kinetic energy
  operator.}
\label{2body}
\end{figure}
%%%%%%%%%%%%%%%%%%%%%%%%%%%%%%%%%%%%%%%%%%%%%%%%%%%%%%%%%%%%%%    
At next-to-leading order (NLO), one includes one perturbative insertion of the
deuteron kinetic term. At next-to-next-to-leading order (NNLO), two insertions
of the kinetic energy enter. At higher orders, terms with more derivatives not
explicitly shown in (\ref{boson_deuteron_lag}) appear. We thus arrive at the
deuteron propagator in the low-energy expansion
\begin{eqnarray}\label{deuteron_prop}
  \bar{\mathcal{D}}(p_0,p)& =&\bar{\mathcal{D}}^{(0)}(p_0,p) +
  \bar{\mathcal{D}}^{(1)}(p_0,p)
  +\bar{\mathcal{D}}^{(2)}(p_0,p)
  +\mathcal{O} (p^3 R^3)\\
  &=&\frac{4\pi}{Mg^2}\left[\frac{1}{-\gamma
  +\sqrt{-Mp_0+\frac{p^2}{4}-\ii\epsilon}}
  -
    \frac{\rho}{2}\frac{\gamma^2+Mp_0-\frac{p^2}{4}}
     {\left(-\gamma+\sqrt{-Mp_0+\frac{p^2}{4}-\ii\epsilon}\right)^2}
   \right.\nonumber\\
  &&\left.
    +\left(\frac{\rho}{2}\right)^2\frac{(\gamma^2+Mp_0-\frac{p^2}{4})^2}
       {\left(-\gamma+\sqrt{-Mp_0+\frac{p^2}{4}-\ii\epsilon}\right)^3} 
       \right]+{\mathcal O} (p^3 R^3)\nonumber\\
  &\equiv& \frac{4\pi}{Mg^2} \mathcal{D} (p_0,p)\;\;,\nonumber
\end{eqnarray}
where $\rho$ is the effective range, defined by performing the effective range
expansion around the bound state. The effective range is assumed to be of
natural size, $\rho \sim R$. We also defined the rescaled propagator
$\mathcal{D}(p_0,p)$ for future convenience and quote the two particle
propagator when the effective range is summed to all orders, but the other
corrections in the effective range expansion are left out:
\begin{equation}
  \label{eq:deuteron_prop_allorders}
  \mathcal{D}(p_0,p)\to\frac{1}{-\gamma+\sqrt{-Mp_0+\frac{p^2}{4}-\ii\epsilon} 
    +\frac{\rho}{2}\left(\gamma^2+Mp_0-\frac{p^2}{4}\right)}\;\;.
\end{equation}
The two particle scattering amplitude is obtained from the $d$ field
propagator by multiplying it by $g^2$. After that, we recognise the familiar
form of the scattering amplitude in the effective range expansion, which had
allowed us to determine the low-energy constants $C_0, C_2$ in terms of
$\gamma$ and $\rho$ in (\ref{deuteron_prop}/\ref{eq:deuteron_prop_allorders}).

Similarly, the leading order three-particle amplitude includes all diagrams
build up of the leading two-body interactions, i.e.~the ones proportional to
$C_0$ in (\ref{boson_lag}) or $g$ and $\Delta$ in (\ref{boson_deuteron_lag}).
The NLO amplitude includes the graphs with one insertion of the deuteron
kinetic energy, the NNLO amplitude includes those with two deuteron kinetic
energy insertions, and at higher orders new interactions not explicitly shown
in (\ref{boson_lag}/\ref{boson_deuteron_lag}) appear. But that is not all.
There also appear graphs containing the three-body interactions proportional
to $D_0,D_2$ in (\ref{boson_lag}) or $\Omega$ and the triton kinetic energy in
(\ref{boson_deuteron_lag}). It has been established that the leading three
body force, i.e.~the one proportional to $D_0$ in (\ref{boson_lag}) or
$\Omega$ in (\ref{boson_deuteron_lag}) appears at LO and
NLO~\cite{3stooges_boson,3stooges_doublet,doubletNLO}. We will re-derive this
fact in a slightly modified way which permits us also to find the proper
generalisation to all orders in the expansion.

Our argument proceeds as follows. We include a specific three-body force term
at a given order $n$ of the expansion \emph{if and only if} that term is
needed to cancel cutoff dependences in the observables which are stronger than
cutoff dependence from the suppressed terms of order $(Q R)^{n+1}$.  The
rationale we follow is thus simply a consequence of the fact that physical
observables cannot depend on the arbitrary value of the cutoff and, from the
point of view of the EFT, cutoff dependences coming from loops and from the
values of the low-energy constants cancel order by order in the expansion.

Before we implement this idea, let us make a few comments about the economy of
computing diagrams at high orders. There is an infinite number of diagrams
contributing at leading order, as shown in Fig.~\ref{lo3bodyeq}.
%%%%%%%%%%%%%%%%%%%%%%%%%%%%%%%%%%%%%%%%%%%%%%%%%%%%%%%%%%
\begin{figure}[!htb]
\begin{center}
  \includegraphics[width=0.9\linewidth,angle=0,clip=true]{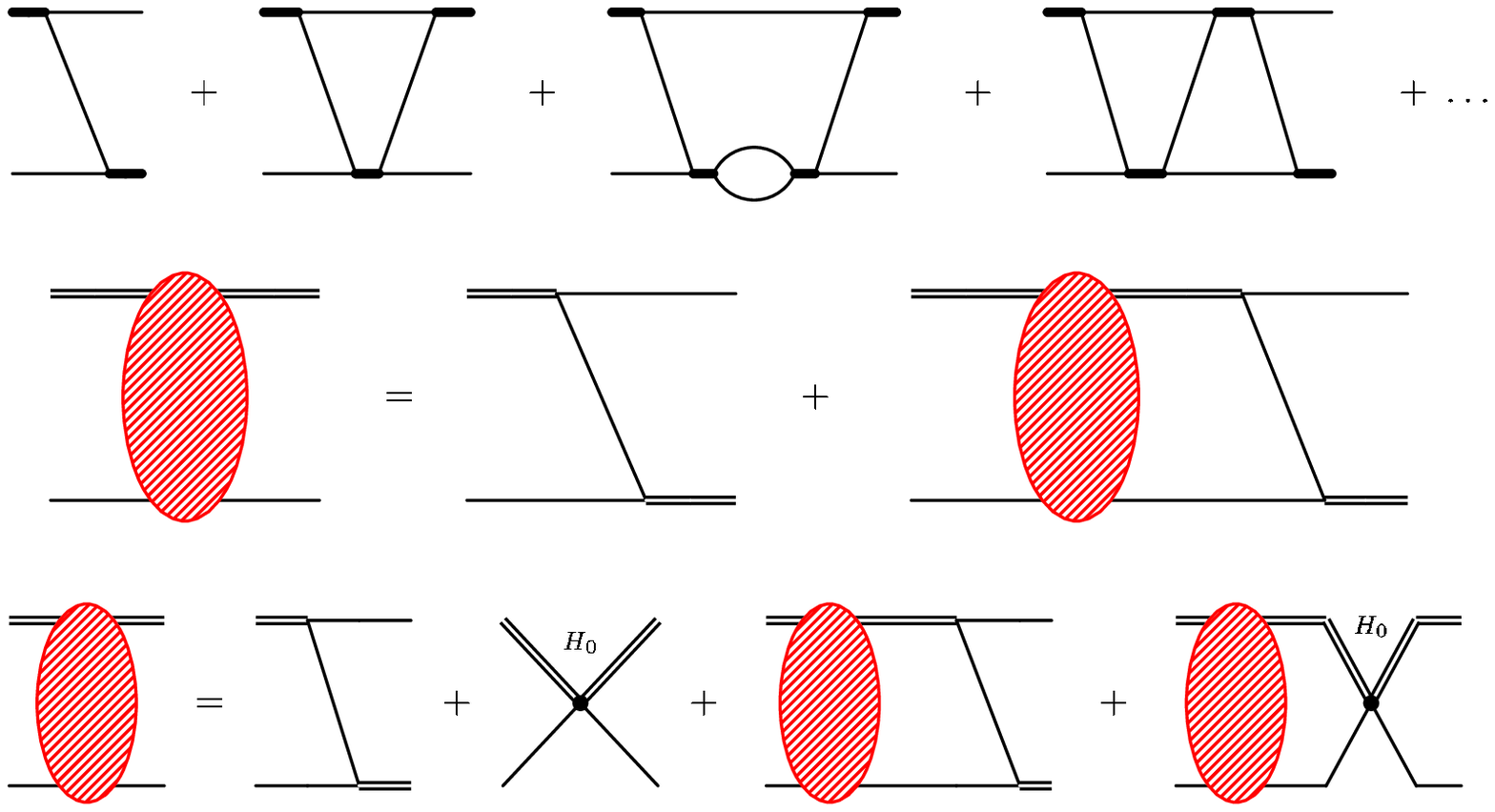}
\end{center}
\vspace*{-0pt}
\caption{Re-summation of the na{\"{\i}}ve LO three-body equation, without
  three-body interactions (top), into the corresponding Faddeev integral
  equation (mid) for the Boson case. Bottom: The same including three-body
  interactions with strength $H_0(\Lambda)$, denoted by the dot. Remaining
  notation as in Fig.~\ref{2body}.}
\label{lo3bodyeq}
\end{figure}
%%%%%%%%%%%%%%%%%%%%%%%%%%%%%%%%%%%%%%%%%%%%%%%%%%%%%%%%%%%%%%    
We sum them up by solving the Faddeev integral equation for point-like
interactions.  In the case where only two-body interactions are present, this
equation was first derived in~\cite{skorny}.  This re-summation is
\emph{necessary} since -- according to the power counting -- all these
diagrams contribute equally to the final amplitude.  The NLO contribution can
then be obtained by perturbing around the LO solution with the deuteron
kinetic energy operator, as shown diagrammatically in
Fig.~\ref{range_correction}.
%%%%%%%%%%%%%%%%%%%%%%%%%%%%%%%%%%%%%%%%%%%%%%%%%%%%%%%%%%
\begin{figure}[!htb]
\begin{center}
  \includegraphics[width=0.7\linewidth,angle=0,clip=true]{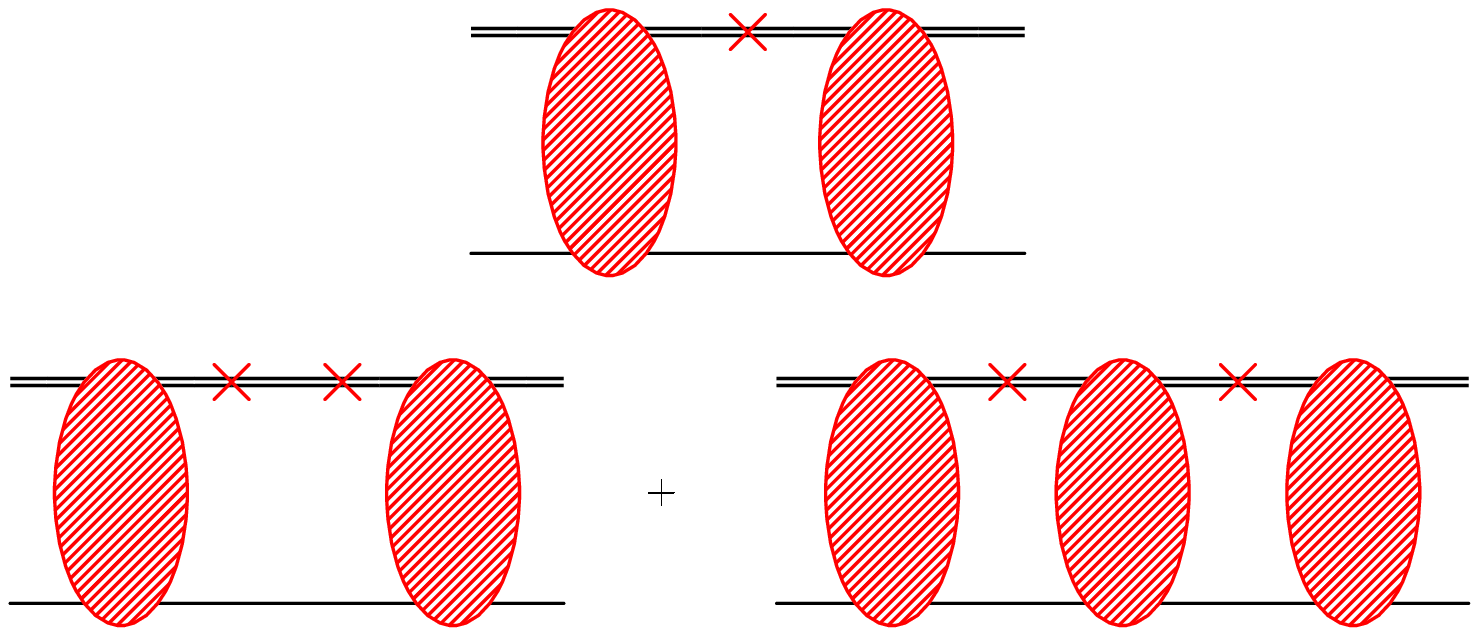}
\end{center}
\vspace*{-0pt}
\caption{The graphs to be computed when inserting the higher order
  corrections to the deuteron propagator perturbatively. Top: NLO. Bottom:
  NNLO. The bottom right diagram makes the knowledge of the full off-shell LO
  amplitude mandatory. Notation as in Fig.~\ref{2body}.}
\label{range_correction}
\end{figure}
%%%%%%%%%%%%%%%%%%%%%%%%%%%%%%%%%%%%%%%%%%%%%%%%%%%%%%%%%%%%%%    
This approach was taken before and is easily implemented~\cite{doubletNLO}.
However, it suffers a drawback at higher orders because one must then insert
the kinetic energy operator two or more times. This is very cumbersome to do
numerically, particularly because of the need to compute the full off-shell LO
amplitude before inserting the deuteron kinetic energy terms, see
Fig.~\ref{range_correction}. A more practical method would be to re-sum all
range corrections into the deuteron propagator and solve a modified integral
equation. This approach has its own drawbacks, including the existence of
spurious bound states in the re-summed deuteron propagator, and was shown to
fail miserably for the phase shifts~\cite{gabbiani}.

We choose a middle ground: First, expand the kernel of the integral equation
perturbatively, and then iterate it by inserting it into the integral
equation. That means, we perform a partial re-summation of the range effects:
We use the deuteron propagator in (\ref{deuteron_prop}) in the integral
equation, which is then solved numerically.
Figure~\ref{range_correction_presum} shows a graphical representation of the
NLO version. This procedures includes some graphs of higher order, for
instance, the NLO calculation includes diagrams like the bottom right one in
Fig.~\ref{range_correction}, but not diagrams like the bottom left one in the
same Figure.  The partial re-summation arbitrarily includes higher order
graphs and consequently does not improve the precision of the calculation. But
the additional diagrams are small in a well-behaved expansion, so the
precision is not compromised, either. On the other hand, this prescription is
much easier to handle numerically than the two other methods. In a nutshell,
the only \emph{necessary} re-summation is the one present at LO. The partial
re-summation of range effects is made only for \emph{convenience}.

%%%%%%%%%%%%%%%%%%%%%%%%%%%%%%%%%%%%%%%%%%%%%%%%%%%%%%%%%%
\begin{figure}[!htb]
\begin{center}
  \includegraphics[width=0.9\linewidth,angle=0,clip=true]{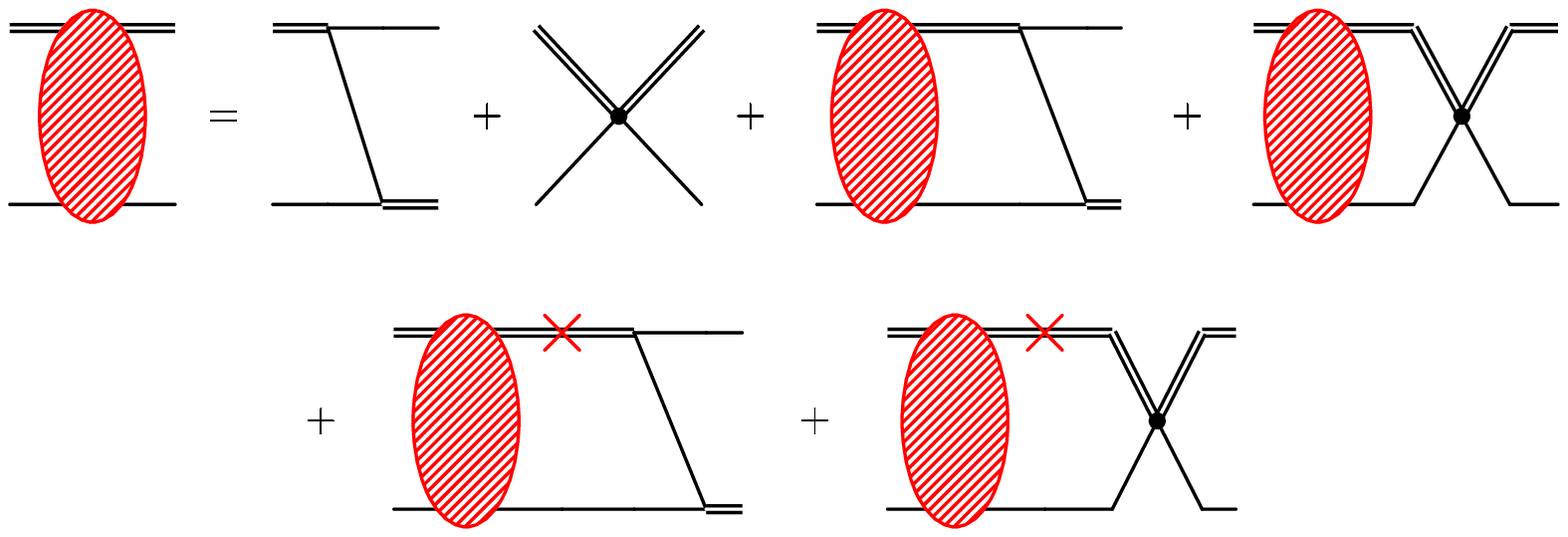}
\end{center}
\vspace*{-0pt}
\caption{The integral equation for the boson case when the kernel is expanded
  to NLO. Notation as in Figs.~\ref{2body} and~\ref{lo3bodyeq}.}
\label{range_correction_presum}
\end{figure}
%%%%%%%%%%%%%%%%%%%%%%%%%%%%%%%%%%%%%%%%%%%%%%%%%%%%%%%%%%

The integral equation describing particle-deuteron scattering one obtains by
including the re-summation discussed above is hence a minor modification of
the one derived in~\cite{skorny,3stooges_boson}. Its NLO version is
graphically represented in Fig.~\ref{range_correction_presum}. The integral
equation for the half off-shell amplitude up to and including NNLO reads:
\begin{equation}\label{integral_eq_boson}
  t_\Lambda(p,k) = \mathcal{K}(p,k)
+\mathcal{H}(p,\Lambda)+\frac{2}{\pi}
 \int\limits_0^\Lambda \dd q\; q^2
  \mathcal{D}(E-\frac{q^2}{2M},q)
  \left[
\mathcal{K}(p,q)+\mathcal{H}(E,\Lambda)\right]
t_\Lambda(q,k)\;\;,
\end{equation}
where
\begin{eqnarray}
\mathcal{H}(E,\Lambda)&=&\mathcal{H}_0(\Lambda)+\mathcal{H}_2(E,\Lambda)\;\;,\\
\mathcal{H}_0(\Lambda)&=&\frac{2
H_0(\Lambda)}{\Lambda^2}=-\frac{\omega^2}{4Mg^2
\Omega}\;\;,\\
\mathcal{H}_2(E,\Lambda)&=&\frac{2H_2(\Lambda)}{\Lambda^4}
\;(ME+\gamma^2)
=\frac{\omega^2}{4Mg^2 \Omega^2}\;(ME+\gamma^2)
\;\;,\\
\mathcal{K}(p,q)&=&
\frac{1}{pq}\ln\left(\frac{p^2+pq+q^2-ME}{p^2-pq+q^2-ME}\right)\;\;,
\end{eqnarray}
and $E=3k^2/(4M)-\gamma^2/M$ is the total centre-of-mass energy.  The on-shell
scattering amplitude $T(k)$ is given by $Z t_\Lambda(k,k)$ , where the
deuteron wave function renormalisation factor is given by
\begin{equation}
  \label{eq:wavefurenbosons}
  Z=\frac{8\pi\gamma}{M}\left(1+\gamma\rho+(\gamma\rho)^2+\dots\right)\;\;.
\end{equation}and is  related to the phase shift $\delta$ in this channel 
by the usual formula
\begin{equation}
  T(k)=Z t_\Lambda(k,k)=\frac{3\pi}{M}\frac{1}{k \cot\delta-\ii k}\;\;.
\end{equation}

As (\ref{integral_eq_boson}) contains the cutoff $\Lambda$, the solution
$t_\Lambda(q)$ for arbitrary $H_0,H_2$ depends in general on the cutoff, too.
Since the low-energy amplitude can not depend on the cutoff, any explicit
dependence on the cutoff has to be cancelled by the cutoff dependence implicit
in $H_0,H_2$, etc. We determine the cutoff dependence of the three-body forces
by imposing this condition order by order in $1/\Lambda$. In this process, we
also determine which three-body force appears at every order of the expansion.

Let us first consider the effect of changing the cutoff from $\Lambda$ to
$\Lambda'\sim \Lambda$. The equation satisfied by the amplitude $t_{\Lambda'}$
analogous to (\ref{integral_eq_boson}) can be written as follows:
\begin{eqnarray}\label{integral_eq_boson_lambdaprime}
  \int\limits_0^\Lambda\! \dd q\;
\underbrace{\Big(\delta(p-q)}_{\sim 1/Q^2}-
\frac{2}{\pi} \underbrace{q^2
  \mathcal{D}(E-\frac{q^2}{2M},q)
   \mathcal{K}(p,q)\Big)}_{\sim 1/Q^2} 
t_{\Lambda'}(q)=\underbrace{\mathcal{K}(p,k)}_{\sim
1/Q^2} 
+\underbrace{\mathcal{H}(E,\Lambda)}_{\sim
1/\Lambda^2}\nonumber\\
  + \frac{2}{\pi}\underbrace{\int\limits_0^\Lambda \dd q\; 
  q^2 {\mathcal
D}(q)\mathcal{H}(E,\Lambda)t_{\Lambda'}(q)}_{\sim
1/(Q\Lambda)}\nonumber\\
+\frac{2}{\pi}
 \underbrace{\int\limits_0^\Lambda \dd q\; q^2
  \mathcal{D}(E-\frac{q^2}{2M},q)
(\mathcal{H}(E,\Lambda')-\mathcal{H}(E,\Lambda))
t_{\Lambda'}(q)}_{\sim 1/(Q\Lambda)}\nonumber\\
  +\underbrace{\mathcal{H}(E,\Lambda')}_{\sim
1/\Lambda^2}
-\underbrace{\mathcal{H}(E,\Lambda)}_{\sim
1/\Lambda^2}\nonumber\\
  +\frac{2}{\pi}
 \int\limits_\Lambda^{\Lambda'} \dd q \;q^2
  \mathcal{D}(E-\frac{q^2}{2M},q)
  \underbrace{\Big(  \mathcal{K}(p,q)}_{\sim
1/(Q\Lambda)}
+\underbrace{\mathcal{H}(E,\Lambda')\Big)}_{\sim
1/(Q\Lambda)}
  t_{\Lambda'}(q)
\end{eqnarray}
The point of this rewriting is to show explicitly as in the last three lines
the difference between the equations satisfied by $t_\Lambda(p)$ and
$t_{\Lambda'}(p)$.  Now we take the cutoffs $\Lambda$ and $\Lambda'$ to be
below the short distance scale $Q\ll \Lambda \ll 1/r \sim 1/R$, but higher
than the infrared scales $Q\sim \gamma\sim k\sim p$.  The asymptotic behaviour
of $t(q)$ is known to be of the form $t(p \gg Q) \sim \sin(s_0 \ln
p/\bar{\Lambda})/(Qp)$~\cite{faddeev} (see Appendix), so $t(p\sim Q)\sim
1/Q^2$ and $t(p\sim \Lambda)\sim 1/(Q\Lambda)$.  As one knows the asymptotic
behaviour of $t_\Lambda(p)$ and assumes $H_0\sim H_2\sim 1$, one can estimate
the size of each term in (\ref{integral_eq_boson_lambdaprime}) as shown in
under-braces. Here, loop momenta $q$ count as $Q$ in ultraviolet finite
integrals but count as $\Lambda$ in divergent integrals, since the integrals
are dominated by momenta of the order $Q$ and $\Lambda$, respectively.

The estimate suggests that the terms on the right hand side describing the
contribution from loop momenta between $\Lambda$ and $\Lambda'$ as well as the
contribution from the three-body forces are small compared to the leading ones
on the left hand side, the former being suppressed by at least one power of
$Q/\Lambda$. Therefore, one na{\ia}vely expects them to be negligible at
leading order. If one indeed discards them, the resulting equation for
$t_{\Lambda'}(p)$ becomes identical to (\ref{integral_eq_boson}) and one has
to conclude that $t_{\Lambda'}(p)\simeq
t_{\Lambda}(p)+\mathcal{O}(Q/\Lambda)$.

The argument above is, however, incorrect. This is due to the fact that left
hand side of (\ref{integral_eq_boson_lambdaprime}) is singular in the limit
$\Lambda\rightarrow\infty$.  More precisely, in this limit, there is a zero
eigenvalue for the operator on the left hand side of
(\ref{integral_eq_boson_lambdaprime}) acting on $t_\Lambda'(p)$:
\begin{equation}\label{zeromode}
\int\limits_0^\infty\! \dd q\; \left(\delta(p-q)- \frac{2}{\pi}
q^2 \mathcal{D}(E-\frac{q^2}{2M},q) \mathcal{K}(p,q)\right)
  a(p)=0\;\; .
\end{equation}
We refer the reader to Ref.~\cite{danilov} which proves this statement
rigorously and also makes it clear that the spectrum of the operator in
(\ref{zeromode}) is continuous. One important point to notice is that the
existence of the zero mode is an ultraviolet phenomenon, independent of the
infrared scales like $E$ or $\gamma$.  In the simpler case $\gamma=0$, the
eigenfunction corresponding to an eigenvalue $\lambda$ can be expressed
analytically at LO as
\begin{equation}\label{asym}
a_\lambda(p)=\sin\left( -\ii s \ln\left(\sqrt{\frac{-3p^2}{4ME}}
+\sqrt{1-\frac{3p^2}{4ME}}\right)  \right)
\end{equation}
with eigenvalue
\begin{equation}\label{eigenvalue}
\lambda=1-\frac{8}{\sqrt{3} s} \frac{\sin(\pi
s/6)}{\cos(\pi s/2)}
\end{equation}
and $-1 < \mathrm{Re}(s) < 1$ The zero mode $a_0(p)$ corresponds to
$s(\lambda=0)=\ii s_0=\ii\; 1.00624\dots$.  The existence of the zero mode
shows that for $\Lambda\rightarrow\infty$, the operator on the left hand side
of (\ref{integral_eq_boson}) has no inverse.  Therefore, the solution to the
integral equation is not unique since one can always add $a_0(p)$ to a given
solution and obtain another valid solution.

For finite $\Lambda$, general theorems guarantee that the spectrum of the
operator is discrete~\cite{kolmogorov_fomin}.  Eq.~(\ref{integral_eq_boson})
then defines the amplitude uniquely. However, as $\Lambda$ is increased the
eigenvalues get closer to each other and, for a generic value of the cutoff,
the eigenvalue closest to zero is of the order $\mathcal{O}(Q/\Lambda)$. Since
the operator in the left hand side of (\ref{integral_eq_boson}) is nearly
singular, terms of order $\mathcal{O}(Q/\Lambda)$ on the right hand side of
(\ref{integral_eq_boson_lambdaprime}) can have an effect on $t_\Lambda(p)$ of
order $\mathcal{O}(1)$.  This ``hypersensitivity'' to cutoff dependences is at
the heart of the apparently paradoxical fact that the solution of
(\ref{integral_eq_boson}) with $\mathcal{H}=0$ is cutoff dependent, even
though the contribution of the high momentum modes are suppressed by
$1/\Lambda$ (that is, are ultraviolet finite).  In general, for the amplitude
to be cutoff independent up to terms of order $\mathcal{O}((Q/\Lambda)^n)$ ,
the right hand side of (\ref{integral_eq_boson_lambdaprime}) has to be cutoff
independent up to terms of order $\mathcal{O}((Q/\Lambda)^{n+1})$ , i.e.~one
order higher than in the standard cases. Notice that the unexpected dependence
of the low-energy amplitude on high momenta is due to two factors: the harder
asymptotic behaviour of $t(p)\sim 1/p$ compared to the perturbation theory
expectation $t(p)\sim 1/p^2$, and the fact that the leading terms in
(\ref{integral_eq_boson_lambdaprime}) form a nearly singular equation that
amplifies any remaining term by one power of $\Lambda/Q$.

We now determine the three-body forces by demanding order by order in
$Q/\Lambda$
\begin{equation}\label{master_renorm}
  {\mathcal H}(E,\Lambda) + \frac{2}{\pi}\int\limits^\Lambda \dd q\; q^2
  \mathcal{D} (E-\frac{q^2}{2M},q)
  \left[ {\mathcal K}(q,p) + {\mathcal H}(E,\Lambda)
  \right]t_\Lambda(q)=\mathrm{const.}\;\;,
\end{equation}
i.e.~the amplitude must be cutoff independent. This (possibly energy
dependent) constant can without loss of generality be set to zero by absorbing
it into a re-defined three body force
\begin{equation}
  \mathcal{H}(E,\Lambda)\to\mathcal{H}(E,\Lambda)-
  \frac{\mathrm{const.}}{1+\displaystyle
    \frac{2}{\pi}\int\limits^\Lambda \dd q\; q^2
  \mathcal{D} (E-\frac{q^2}{2M},q)\;t_\Lambda(q)}\;\;.
\end{equation}
Imposing (\ref{master_renorm}) with $\mathrm{const.}=0$ order by order in the
low energy expansion, we can determine which three-body forces are required at
any given order, and how they depend on the cutoff. As we are interested only
in the UV behaviour, i.e.~in the asymptotics, and no IR divergences occur, we
can safely neglect the IR limit of the integral.

At leading order, $\gamma\ll\Lambda$, and we can approximate
\begin{equation}\label{LO_approx}
\begin{split}
  \mathcal{D}(E-\frac{q^2}{2M},q) &\simeq \sqrt{\frac{4}{3} }
  \frac{1}{q}+\dots\;\;,\\
  {\mathcal K}(q,p) &\simeq \frac{2}{q^2}+\dots\\
  {\mathcal H}(E,\Lambda) &\simeq \frac{2
    H_0(\Lambda)}{\Lambda^2}+\dots\\
  t(q) &\simeq A(k/\Lambda,\gamma/\Lambda) \frac{\sin[s_0
    \ln(q/\bar\Lambda)]}{q}+\dots\;\;.
\end{split}
\end{equation}
where a normalisation factor $A$ was included into the asymptotic form of
$t(q)$. The essential observation is that the terms of order
$\mathcal{O}(Q/\Lambda)$ are independent of $E$ (or $k$) and hence can be made
to vanish by adjusting only $H_0$ and not $H_2$ or any of the higher
derivative three-body forces. In fact, using the substitutions of
(\ref{LO_approx}) in (\ref{master_renorm}) and neglecting higher order
corrections for naturally-sized $H_0$, we find
\begin{equation}
  \label{HLO}
H(\Lambda) = \frac{\sin[s_0
\ln\frac{\Lambda}{\bar{\Lambda}} + \arctan s_0]}
{\sin[s_0 \ln\frac{\Lambda}{\bar{\Lambda}} -
\arctan s_0]}
\end{equation}
as obtained previously~\cite{3stooges_boson}.

At NLO, we impose (\ref{master_renorm}) to order $\mathcal{O}(1/\Lambda^2)$.
Thus, we expand one order higher than (\ref{LO_approx})
\begin{equation}\label{NLO_approx}
\begin{split}
  \mathcal{D}(E-\frac{q^2}{2M},q) &\simeq \sqrt{\frac{4}{3}}\frac{1}{q}+
  \frac{4}{3}\frac{\gamma}{q^2}+\frac{r}{2}\dots\ 
  \\
  {\mathcal K}(q,p) &\simeq \frac{2}{q^2}+\dots\\
  {\mathcal H}(E,\Lambda) &\simeq \frac{2 H_0(\Lambda)}{\Lambda^2}+\dots\;\;.
\end{split}
\end{equation}
The sub-leading asymptotics at NLO depends on whether $\Lambda\ll 1/r$ or
$\Lambda\gg 1/r$.  In the first case, there is only a perturbative change from
the LO asymptotics, and we can find it analytically (see Appendix)
\begin{equation}\label{NLO_approx_t}
\begin{split}
  t(q) \simeq A(k/\Lambda) \Big[ \frac{\sin[s_0 \ln(q/\bar\Lambda)]}{q}&+
  |B_1| \;r\sin[s_0 \ln(q/\bar\Lambda)+\arg(B_1)]\\
  &+|B_{-1}| \;8 \gamma \frac{\sin[s_0
    \ln(q/\bar\Lambda)+\arg(B_{-1})]}{q^2}+\dots\Big]\;\;.
\end{split}
\end{equation}
Again we see that, to this order, no terms dependent on $E$ or $k$ are
present.  Only the $H_0(\Lambda)$ term is necessary for renormalisation.
Plugging (\ref{NLO_approx}/\ref{NLO_approx_t}) back in (\ref{master_renorm})
and keeping only the terms of order $1/\Lambda^2$, one can derive the
sub-leading correction to (\ref{HLO}), but the expression is long and of no
immediate use.

A new three-body force enters at NNLO. We require now that the terms of order
$(Q/\Lambda)^3$ on the right hand side of
(\ref{integral_eq_boson_lambdaprime}) should vanish, which guarantees that the
amplitude is cutoff independent up to terms of order $(Q/\Lambda)^2$. At this
order though, there are terms proportional to $ME$ arising from expanding the
kernel in powers of $q/\Lambda$ and $\sqrt{ME}/\Lambda$. Indeed, such
corrections appear only at even orders.  Only a three-body force term which
also contains a dependence on the external momenta $k^2$ and $ME$ (both
equivalent on-shell) can absorb them. That means that $H_2$ appears at NNLO.
In addition, there is a large number of momentum independent terms that
renormalise $H_0$ and provide the NNLO corrections to (\ref{HLO}).  The same
pattern repeats at higher orders.  Every two orders, a new three-body force
term appears with two extra derivatives, guaranteeing that the new three-body
forces are analytic in $ME\propto k^2$ on shell.

If we choose $\Lambda \agt 1/r$, the argument is changed only by the change in
the asymptotic form of $t(p)$. This affects the explicit forms for the running
of $H_0,H_2$, but not the fact that they are necessary and sufficient to
perform the renormalisation.

%%%%%%%%%%%%%%
\section{Three nucleons in the triton channel}
\label{sec:triton}

It is not difficult to include the spin and isospin degrees of freedom and
generalise the discussion in the previous section to the case of the
three-nucleon system. As discussed elsewhere, the $^2\mathrm{S}_{\frac{1}{2}}$
channel -- to which $^3$He and $^3$H belong -- is qualitatively different from
the other three-nucleon channels.  This difference can be traced back to the
effect of the exclusion principle and the angular momentum repulsion barrier.
In all the other channels, it is either the Pauli principle or an angular
momentum barrier (or both) which forbids the three particles to occupy the
same point in space.  As a consequence, the kernel describing the interaction
among the three nucleons is, unlike in the bosonic case, repulsive in these
channels.  The zero mode of the bosonic case, discussed above, describes a
bound state since it is a solution of the homogeneous version of the Faddeev
equation. As such, it is not expected to appear in the case of repulsive
kernels and, in fact, it does not.  For the $^3$He-$^3$H channel however, the
kernel is attractive and, as we will see below, closely related to the one in
the bosonic case.  Consequently, the arguments in the previous section are
only slightly modified for the case of three nucleons in the triton channel.

The three-nucleon Lagrangean analogous to (\ref{boson_deuteron_lag}) is given
by
\begin{equation}\label{triton_deuteron_lag}
\begin{split}
  {\mathscr L}&=N^\dagger\left(\ii\partial_0+\frac{\nabla^2}{2M}\right)N
  +d_s^{A\dagger}\left(-\ii\partial_0-\frac{\nabla^2}{4M}+\Delta_s\right)d_s^A
  +d_t^{i\dagger}\left(-\ii\partial_0-\frac{\nabla^2}{4M}+\Delta_t\right)
  d_t^i\\
  &+t^\dagger\left(\ii\partial_0+\frac{\nabla^2}{6M}+\frac{\gamma^2}{M}+
    \Omega\right)t -g_s\left( d_s^{A \dagger} (N^T P^A N) +\text{H.c.}
  \right) -g_t\left( d_t^{i \dagger} (N^T P^i N) +\text{H.c.}
  \right)\\
  &-\omega_s \left( t^\dagger (\tau^A N) d_s^{A} +\text{H.c.} \right)
  -\omega_t \left( t^\dagger (\sigma^iN) d_t^{i} +\text{H.c.} \right)
  +\dots\;\;,
\end{split}
\end{equation}
where $N$ is the nucleon iso-doublet and the auxiliary fields $t$, $d_s^A$ and
$d_t^i$ carry the quantum numbers of the $^3$He-$^3$H spin and isospin
doublet, $^1S_0$ di-nucleon and the deuteron, respectively.  The projectors
$P^i$ and $P^A$ are defined by
\begin{equation}
\begin{split}
  P^i=\frac{1}{\sqrt{8}}\tau_2 \sigma_2\sigma^i\\
  P^A=\frac{1}{\sqrt{8}}\sigma_2 \tau_2\tau^A \;\;,
\end{split}
\end{equation}
where $A=1,2,3$ and $i=1,2,3$ are iso-triplet and vector indices and $\tau^A$
($\sigma^i$) are isospin (spin) Pauli matrices.

When the auxiliary fields are integrated out of (\ref{triton_deuteron_lag}),
three different kinds of three-body force terms without derivatives seem to be
generated:
\begin{equation}\label{LO_3bodyforce}
\begin{split}
  {\mathscr L}_\mathrm{3 body}&=(N^T P^iN)^\dagger
  (N^\dagger \sigma^i \sigma^iN)( N^T P^iN)\\
  &=(N^T P^AN)^\dagger (N^\dagger \tau^A \tau^AN) (N^T
  P^AN)\\
  &=-(N^T P^AN)^\dagger (N^\dagger \tau^A
  \sigma^iN) (N^T P^iN)\\
  &\propto (N^\dagger N)^3\;\;.
\end{split}
\end{equation}
This last form for the three-body force is explicitly
Wigner-$SU(4)$-symmetric, i.e.~it is symmetric by a generic unitary
transformation of the four fields corresponding to the $2$(spin) $\times
2$(isospin)$=4$ nucleon states~\cite{su4} and corresponds to the singlet and
triplet effective range parameters to be identical.  As observed before, the
term in (\ref{LO_3bodyforce}) is the only possible three-body force without
derivatives.  Thus, the different couplings $\omega_s$,$\omega_t$ are
redundant and we can take them to be determined by only one free parameter
\begin{equation}
\frac{H_0}{\Lambda^2}=-\frac{6}{Mg_t^2}\frac{\omega_t^2}{\Omega}
=-\frac{6}{Mg_s^2}\frac{\omega_s^2}{\Omega}
=\frac{6}{Mg_sg_t}\frac{\omega_s\omega_t}{\Omega}\;\;.
\end{equation} 
Contrary to the terms without derivatives, there are different, inequivalent
three-body force terms with \emph{two} derivatives. As we will see below,
however, only the $SU(4)$-symmetric combination is enhanced compared to the
na{\ia}ve dimensional analysis estimate and appears at the order we work. We
keep then only this special combination.  This amounts to choosing
\begin{equation}
\frac{H_2}{\Lambda^4}=\frac{6}{M^2g_t^2}\frac{\omega_t^2}{\Omega^2}
=\frac{6}{M^2g_s^2}\frac{\omega_s^2}{\Omega^2}
=-\frac{6}{M^2g_sg_t}\frac{\omega_s\omega_t}{\Omega^2}\;\;.
\end{equation}
With these choices, we have six parameters left at NNLO: $\Delta_s, g_s$
(determined by the scattering length and effective range in the $^1S_0$
two-nucleon channel), $\Delta_t, g_t$ (determined by the deuteron pole and
effective range in the deuteron channel), $H_0$ and $H_2$ (to be determined by
three-nucleon data).

The derivation of the integral equation describing neutron-deuteron scattering
has been discussed before~\cite{skorny,3stooges_doublet}. We present here only
the result, including the new term generated by the two-derivative three-body
force.  Two amplitudes get mixed: $t_s$ describes the $d_t + N\rightarrow d_s
+ N$ process, and $t_t$ describes the $d_t + N\rightarrow d_t + N$ process:
\begin{eqnarray}
% hgrie: Matrix form, with possibly wrong factors
%\lefteqn{\left(
%\begin{array}{c}
%  t_s\\[1.5ex]t_t
%\end{array}
%\right)(p,k)= \frac{1}{4}\left(
%\begin{array}{c}
%  3\mathcal{K}(p,k)+2\mathcal{H}(E,\Lambda)\\[1.5ex]
%  \mathcal{K}(p,k)+2\mathcal{H}(E,\Lambda)
%\end{array}
%\right)+\nonumber}\\[2ex]
%&&+
%\frac{1}{2\pi}
% \int\limits_0^\Lambda \dd q\; q^2 \;
% \left(
%\begin{array}{cc}
%  \mathcal{D}_s(q)\left[\mathcal{K}(p,q)+2\mathcal{H}(E,\Lambda)\right]&
%   \mathcal{D}_t(q)\left[3\mathcal{K}(p,q)+2\mathcal{H}(E,\Lambda)\right]
  %\\[1.5ex]
%     \mathcal{D}_s(q)\left[3\mathcal{K}(p,q)+2\mathcal{H}(E,\Lambda)\right]&
%   \mathcal{D}_t(q)\left[\mathcal{K}(p,q)+2\mathcal{H}(E,\Lambda)\right]
%\end{array}
%\right)
%\;
%\left(
%\begin{array}{c}
%  t_s\\[1.5ex]t_t
%\end{array}
%\right)(q,k)
%***************
 t_s(p,k)& =  \frac{1}{4}\left[3\mathcal{K}(p,k)
+2\mathcal{H}(E,\Lambda)\right]+\dis\frac{1}{2\pi}
 \int\limits_0^\Lambda \dd q\; q^2 
    & \left[\mathcal{D}_s(q)\left[\mathcal{K}(p,q)+2\mathcal{H}(E,\Lambda)
      \right]
t_s(q)\right.\nonumber\\
       &&\left.+\mathcal{D}_t(q)\left[3\mathcal{K}(p,q)+2\mathcal{H}(E,\Lambda)
       \right]
t_t(q)\right] \label{int_equation_triton}\nonumber\\
 t_t(p,k)& = \frac{1}{4}\left[\mathcal{K}(p,k)
+2\mathcal{H}(E,\Lambda)\right]+\dis\frac{1}{2\pi}
 \int\limits_0^\Lambda \dd q\; q^2 
  &\left[ \mathcal{D}_t(q)\left[ 
\mathcal{K}(p,q)+2\mathcal{H}(E,\Lambda)\right]
t_t(q)\right.\nonumber\\
       & & \left.+\mathcal{D}_s(q)
       \left[3\mathcal{K}(p,q)+2\mathcal{H}(E,\Lambda)\right]
t_s(q)\right]\;\;,
\end{eqnarray}
where $\mathcal{D}_{s,t}(q)=\mathcal{D}_{s,t}(E-\frac{q^2}{2M},q)$ are the
propagators analogous to (\ref{deuteron_prop}).  For the spin-triplet
$\mathrm{S}$-wave channel, one replaces the two boson binding momentum
$\gamma$ and effective range $\rho$ by the deuteron binding momentum
$\gamma_t=45.7025\;\mathrm{MeV}$ and effective range
$\rho_t=1.764\;\mathrm{fm}$. Because there is no real bound state in the spin
singlet channel of the two-nucleon system, its free parameters are better
determined by the scattering length $a_s=1/\gamma_s=-23.714\;\mathrm{fm}$ and
the effective range $r_s=2.73\;\mathrm{fm}$ at zero momentum, i.e.~instead of
(\ref{deuteron_prop}),
% one has after resumming effective range effects
%analogous to (\ref{eq:deuteron_prop_allorders})
%\begin{equation}
%  \label{eq:transvestite_prop_allorders}
%  \mathcal{D}_s(q_0,q)\to\frac{1}{-\gamma_s+\sqrt{\frac{3p^2}{4}-Mp_0}+
%    \frac{r_s}{2}\left(Mp_0-\frac{3p^2}{4}\right)}
%\end{equation}
%so that the expansion of $\mathcal{D}_s(q):=\mathcal{D}_s(E-\frac{q^2}{2M},q)$
%to NNLO reads
\begin{eqnarray}
  \label{eq:transvestite_prop}
\mathcal{D}_s(p)&=&
  \frac{1}{-\gamma_s+\sqrt{\frac{3}{4}(p^2-k^2)+\gamma_t^2}}+\\
  &&+
    \frac{3r_s}{8}\frac{p^2-k^2+\frac{4}{3}\gamma_t^2}{
      \left(-\gamma_s+\sqrt{\frac{3}{4}(p^2-k^2)+\gamma_t^2}\right)^2}
    +\left(\frac{3 r_s}{8}\right)^2\frac{(p^2-k^2+\frac{4}{3}\gamma_t^2)^2}{
      \left(-\gamma_s+\sqrt{\frac{3}{4}(p^2-k^2)+\gamma_t^2
        }\right)^3}\nonumber\;\;.
\end{eqnarray}
%and again $\mathcal{D}_s(q):=\mathcal{D}_s(E-\frac{q^2}{2M},q)$.

The neutron-deuteron $J=1/2$ phase shifts $\delta$ is determined by the
on-shell amplitude $t_t(k,k)$, multiplied with the wave function
renormalisation for the deuteron as given in
(\ref{eq:wavefurenbosons})~\footnote{The mixing with the
  $^4\mathrm{D}_{\frac{1}{2}}$ channel vanishes at the order we consider
  here.}
\begin{equation}
T(k)=Z t_t(k,k)=\frac{3\pi}{M}\frac{1}{k \cot\delta-\ii k}\;\;.
\end{equation}
The renormalisation of (\ref{int_equation_triton}) is better understood after
introducing the variables $t_+=t_s+t_t$, $t_-=t_s-t_t$, ${\mathcal
  D}_+=(\mathcal{D}_s+\mathcal{D}_t)/2$ and $\mathcal{D}_-=({\mathcal
  D}_s-\mathcal{D}_t)/2$, in terms of which (\ref{int_equation_triton}) reads
 \begin{eqnarray}
%hgrie: again matrix form, with possibly wrong prefactors   
%  \lefteqn{\left(
%\begin{array}{c}
%  t_+\\[1.5ex]t_-
%\end{array}
%\right)(p,k)= \left(
%\begin{array}{c}
%  \mathcal{K}(p,k)+\mathcal{H}(E,\Lambda)\\[1.5ex]
%  \frac{1}{2}\mathcal{K}(p,k)
%\end{array}
%\right)+\nonumber}\\[2ex]
%&&+
%\frac{2}{\pi}
% \int\limits_0^\Lambda \dd q\; q^2 \;
% \left(
%\begin{array}{cc}
%  \mathcal{D}_+(q)\left[\mathcal{K}(p,q)+\mathcal{H}(E,\Lambda)\right]&
%  \mathcal{D}_-(q)\left[\mathcal{K}(p,q)+\mathcal{H}(E,\Lambda)\right]
%\\[1.5ex]
%     -\frac{1}{4}\mathcal{D}_-(q)\mathcal{K}(p,q)&
%   -\frac{1}{4}\mathcal{D}_+(q)\mathcal{K}(p,q)
%\end{array}
%\right)
%\;
%\left(
%\begin{array}{c}
%  t_+\\[1.5ex]t_-
%\end{array}
%\right)(q,k)
 t_+(p,k)&\!\!\! =\!\!\! & \mathcal{K}(p,k)
+\mathcal{H}(E,\Lambda)\nonumber\\
&&+\frac{2}{\pi}
 \int\limits_0^\Lambda \dd q\;\; q^2     
\;[\mathcal{K}(p,q)+\mathcal{H}(E,\Lambda)]\;
 [\mathcal{D}_+(q)t_+(q)
  +\mathcal{D}_-(q)t_-(q)]
\label{int_equation_t+}\\
 t_-(p,k)& = &\frac{1}{2}\;\mathcal{K}(p,k) - \frac{1}{\pi}
 \int\limits_0^\Lambda \dd q\;\; q^2 
     \mathcal{K}(p,q)\;
[\mathcal{D}_+(q)t_-(q)
 +\mathcal{D}_-(q)t_+(q)]\;\;.
\label{int_equation_t-}
\end{eqnarray}
The essential observation is that in the Wigner-$SU(4)$-limit
($\mathcal{D}_-=0$), the two equations in (\ref{int_equation_triton})
decouple and that, in the ultraviolet r\'egime where the differences between
$\gamma_s$ and $\gamma_t$ disappear, the $SU(4)$ limit is recovered.  The
amplitude $t_+$ satisfies the same equation (\ref{integral_eq_boson}) as the
boson case above, and $t_-$ satisfies the same equation as the
$^4S_\frac{3}{2}$ amplitude~\cite{skorny,3stooges_doublet} which does not
require any renormalisation at this order.  Thus, the arguments given for the
renormalisation in the bosonic case applies with only minor changes for the
$^2\mathrm{S}_\frac{1}{2}$ channel.

Let us now look at these changes in detail. First, we note that the asymptotic
behaviour of $t_-(p,k)$ is driven by the term containing
$\mathcal{D}_-(E-\frac{q^2}{2M},q)t_+(q)$:
\begin{equation}
\begin{split}
  t_-(p,k)&\sim \int\limits_0^\infty \dd q\;\; q^2 {\mathcal K}(p,q)
  \;\frac{\gamma_s-\gamma_t}{q^2} \;\frac{\sin[s_0
    \ln(q/\bar\Lambda)]}{q}\\
  &\sim(\gamma_s-\gamma_t)\;\frac{\sin[s_0 \ln(p/\bar\Lambda)+\arg
    B_{-1}]}{p^2}\;\;.
\end{split}
\end{equation}
We can separate (\ref{int_equation_t+}) and (\ref{int_equation_t-}) as in
(\ref{integral_eq_boson_lambdaprime}), retaining on the left hand side the
piece proportional to $\mathcal{D}_+$. Eq.~(\ref{int_equation_t+}) has the
same zero mode we found in the bosonic case so, just as in the bosonic case,
the amplitude is sensitive to the pieces of order $1/(Q\Lambda)$ on the right
hand side already at leading order.  On the other hand, thanks to a crucial
change of sign in the kernel, (\ref{int_equation_t-}) does \emph{not} have a
zero mode and does not present this enhanced sensitivity on $1/\Lambda$
suppressed terms.  The terms of order $1/\Lambda$ or higher can be estimated
by approximating the kernel in the ultraviolet as in
(\ref{LO_approx}/\ref{NLO_approx})\footnote{For the sake of the simplicity of
  the argument, we take here $\rho_s=r_s$ which is in general true up to
  higher order corrections, in order to use the same effective range expansion
  of singlet and triplet $NN$.}
\begin{equation}\label{approx_triton}
\begin{split}
  \mathcal{D}_+(E-\frac{q^2}{2M},q) &\simeq \frac{1}{2}\left[\frac{4}{\sqrt{3}
      }\frac{1}{q}+
    \frac{4}{3}\frac{\gamma_s+\gamma_t}{q^2}+\frac{\rho_s+\rho_t}{2}
    \dots\right]
  \\
  \mathcal{D}_-(E-\frac{q^2}{2M},q) &\simeq \frac{1}{2}\left[
    \frac{4}{3}\frac{\gamma_s-\gamma_t}{q^2}+\frac{\rho_s-\rho_t}{2}
    \dots\right]\;\;.
\end{split}
\end{equation}
By combining the asymptotic behaviour of $t_+$, $t_-$ and
(\ref{approx_triton}), one can find the contribution of the high momentum
modes ($p\sim\Lambda$) from the different terms in
(\ref{int_equation_t+}/\ref{int_equation_t-}).  Starting from
(\ref{int_equation_t+}), we see that first, $\mathcal{D}_+ t_+$ is of order
$1/(Q\Lambda)$; second, it is identical to its bosonic counterpart; and third,
it is absorbed by the same $H_0$ as in (\ref{HLO}). The same term also
generates a term $\sim 1/\Lambda^2$ (equal to the one in the bosonic case
after the substitution $\gamma\rightarrow(\gamma_s+\gamma_t)/2,
r\rightarrow(\rho_s+\rho_t)/2$ ), and one scaling like $\sim ME/(Q\Lambda^3)$,
absorbed into $H_0$ and $H_2$, respectively, in close analogy to the bosonic
case. The term $\mathcal{D}_- t_-$ contributes with a term of order $\sim
(\gamma_s-\gamma_t)^2/(Q\Lambda^3)$, which is absorbed into $H_0$. As
mentioned before, (\ref{int_equation_t-}) does not exhibit the enhanced
sensitivity to terms suppressed by powers of $1/\Lambda$.  Thus, to guarantee
that $t_-$ is cutoff independent up to terms of order $1/(Q\Lambda^n)$, it is
enough to make (\ref{int_equation_t-}) cutoff independent up to terms of the
\emph{same} order. Both terms $\mathcal{D}_+ t_-$ and $\mathcal{D}_- t_+$
contribute with terms starting at order $(\gamma_s-\gamma_t)/(Q\Lambda^2)$.
However, it is well-known that the only point-like three-body force without
derivatives is necessarily $SU(4)$-symmetric. In the renormalisation group
sense, the hence irrelevant (i.e.~spurious) interaction induced at finite
$\Lambda$ is irrelevant and serves only to repair violations of Wigner's
$SU(4)$ symmetry by our choosing a cutoff, which is equivalent to smearing out
the three-body operator.  In closing, we stress that the approximations in
(\ref{approx_triton}) are performed only in the ultraviolet r\'egime, when
analysing the contribution of order $\sim 1/(Q\Lambda)$ to the integral
equation.  This analysis determines only which three-body forces are needed to
render cutoff invariant results.  The full equation with $\gamma_s \neq
\gamma_t, \rho_s \neq \rho_t$ has to be solved to compute the phase shifts,
and to adjust the three-body forces.

%%%%%%%%%%%%%%
\section{Numerical results for ${}^2\mathrm{S}_\frac{1}{2}$ neutron-deuteron
  scattering} 
\label{sec:numerics}

We numerically solved the coupled equations in (\ref{int_equation_triton}) up
to NNLO using besides the values given above the iso-spin averaged nucleon
mass $M=938.9\;\mathrm{MeV}$.  The computational effort in solving Faddeev
equations for separable potentials, as it is the case here, is trivial, and
the reader is invited to download the Mathematica codes used from {\tt
  http://www-nsdth.lbl.gov}.  To deal with the deuteron pole and the
logarithmic singularity (which is integrable but generates numerical
instabilities), we follow Hetherington and Schick~\cite{HetheringtonSchick} to
first solve the integral equation on a contour in the complex plane and then
use the equation again to find the amplitude on the real axis.  We determine
the two-nucleon parameters from the deuteron binding energy and triplet
effective range (defined by an expansion around the deuteron pole, not at zero
momentum) and the singlet scattering length and effective range (defined by
expanding at zero momentum). Finally, we fix the three-body parameters as
follows: Because we defined $H_2$ such that it does not contribute at zero
momentum scattering, one can first determine $H_0$ from the
${}^2\mathrm{S}_\frac{1}{2}$ scattering length
$a_3=(0.65\pm0.04)\;\mathrm{fm}$~\cite{doublet_sca}.  At LO and NLO, this is
the only three-body force entering, but at NNLO, where we saw that $H_2$ is
required, it is determined by the triton binding energy
$B_3=8.48\;\mathrm{MeV}$. The cutoff is varied as discussed below.

The error in the determination of the ${}^2\mathrm{S}_\frac{1}{2}$ scattering
length of about $7\%$, last measured 1971, dominates our errors at NNLO. A
better experimental determination of this parameter is not only a necessary
condition for the improvement of the accuracy of our method. It is also
important information for fixing any three-body force in sophisticated models
of nuclear physics, given that there is only scarce information on phase
shifts in the triton channel.

%%%%%%%%%%%%%%%%%%%%%%%%%%%%%%%%%%%%%%%%%%%%%%%%%%%%%%%%%%
\begin{figure}[!htb]
\begin{center}
  \includegraphics[width=0.8\linewidth,angle=0,clip=true]{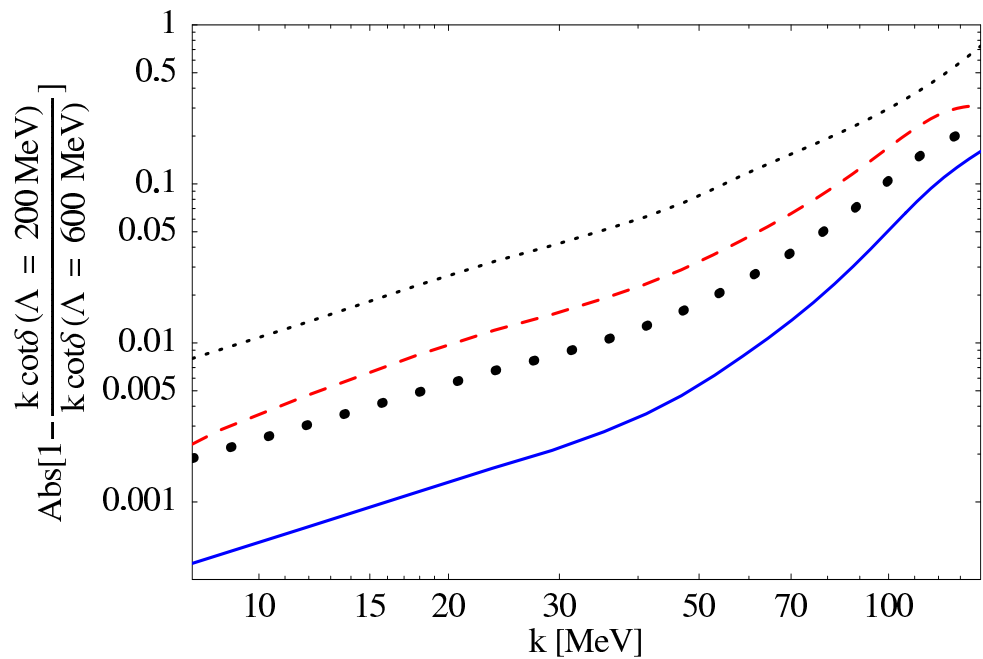}
\end{center}
\vspace*{-0pt}
\caption{Logarithm of the cutoff variation of phase shifts between 
  $\Lambda=200$ MeV and $\Lambda=600$ MeV as function of the centre-of-mass
  momentum.  The short dashed, long dashed and solid line correspond
  respectively to LO, NLO and NNLO. The dots correspond to a ``NNLO''
  calculation without the three-body force $H_2$.}
\label{error}
\end{figure}
%%%%%%%%%%%%%%%%%%%%%%%%%%%%%%%%%%%%%%%%%%%%%%%%%%%%%%%%%%%%%%   

The importance of isospin-violating effects can be estimated to be of higher
order.  In fact, the splitting between the $^1S_0$ scattering lengths between
neutron-neutron ($a_{nn}$) and neutron-proton ($a_{np}$) can be estimated as
$(1/a_{nn} - 1/a_{np})/Q$, where $Q$ is the typical momentum in the system.
Taking $Q \sim \sqrt{M B_3}$ this is a correction of about $2\%$, smaller than
the cutoff variation we see at NNLO.  For future high precision calculations,
however, their inclusion is certainly required.

Unless otherwise stated, all calculations were performed with the cutoff at
$600$ MeV. In the phase shift plots, we include the range of the NNLO result
as the cutoff is varied between $200$ and $600$ MeV. Albeit it neglects other
sources of error at low cutoffs, this is a reasonable estimate of the errors,
especially at high momentum. A definite answer for the size of the errors,
however, can only be given by a higher-order calculation. The lower limit of
this variation was chosen to be of the order of the pion mass. The upper limit
is rather arbitrary since beyond $\Lambda\approx 400$ MeV, the phase shifts
are essentially cutoff independent.  The phase shifts are cutoff independent
up to the order the calculation is valid. This is demonstrated in the plot in
Fig.~\ref{error}, where we show the relative difference in $k \cot \delta$
caused by varying the cutoff between $\Lambda=200$ MeV and $\Lambda=600$ MeV
at different orders.  Two points are worth noticing.  First, the cutoff
variation decreases steadily as we increase the order of the calculation and
is of the order of $(k/\Lambda)^n, (\gamma/\Lambda)^n$, where $n$ is the order
of the calculation and $\Lambda=200$ MeV is the smallest cutoff used.  Second,
the errors increase with increasing momentum, as one would expect from the
fact that there are errors of order $(k/\Lambda)^n$. The slope does not change
significantly as we go to higher orders due to the $ (\gamma/\Lambda)^n$
terms. We also show in the same figure the same cutoff dependence of phase
shifts but now calculated at NNLO \emph{without} the three-body force $H_2$.
It shows that not much improvement is obtained over the cutoff dependence of
the NLO calculation, lending credence to the power counting discussed above.
%%%%%%%%%%%%%%%%%%%%%%%%%%%%%%%%%%%%%%%%%%%%%%%%%%%%%%%%%%%%%%%%%%
\begin{figure}[!htb]
  \begin{center}
    \includegraphics*[width=0.9\linewidth, clip=true]{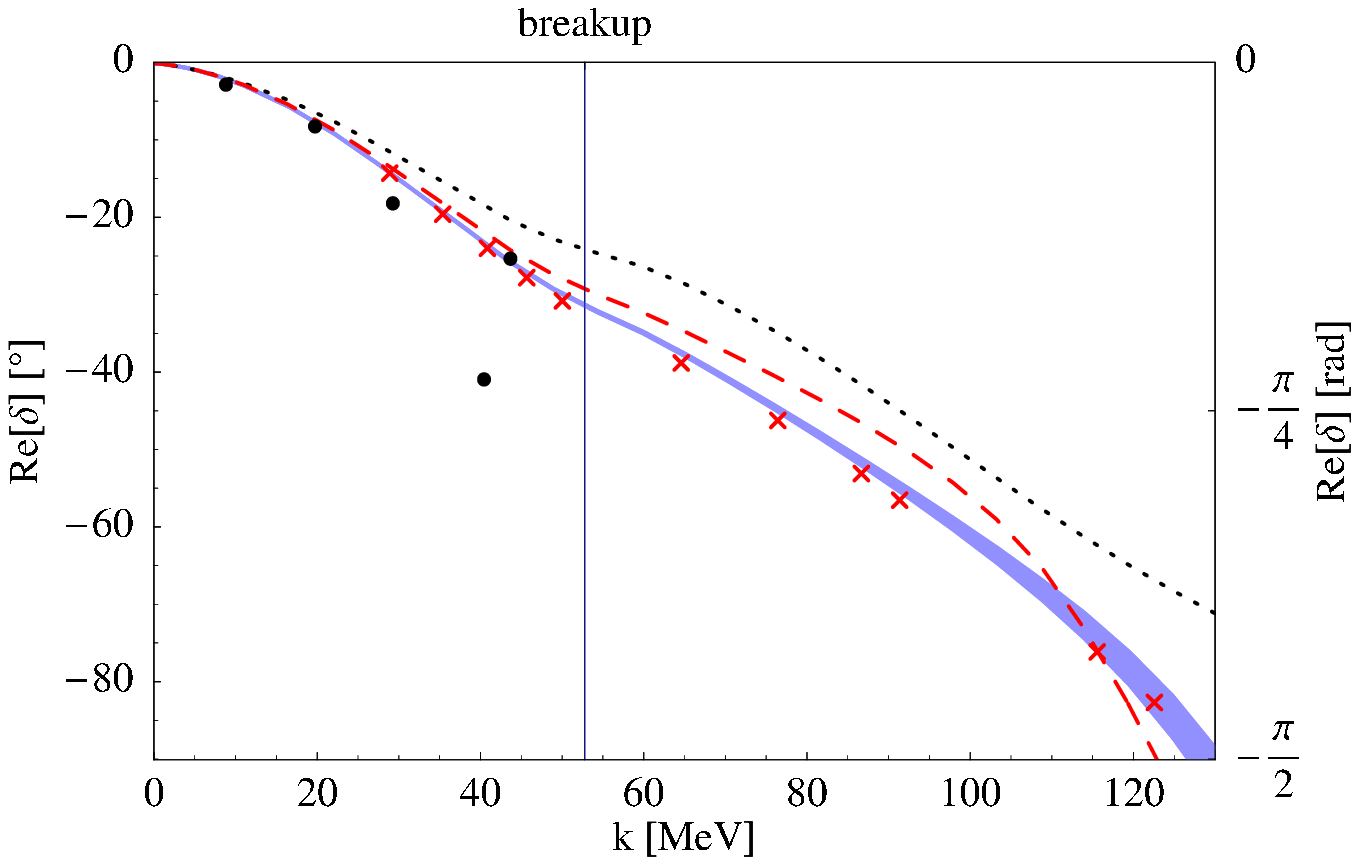}
    \includegraphics*[width=0.86\linewidth, clip=true]{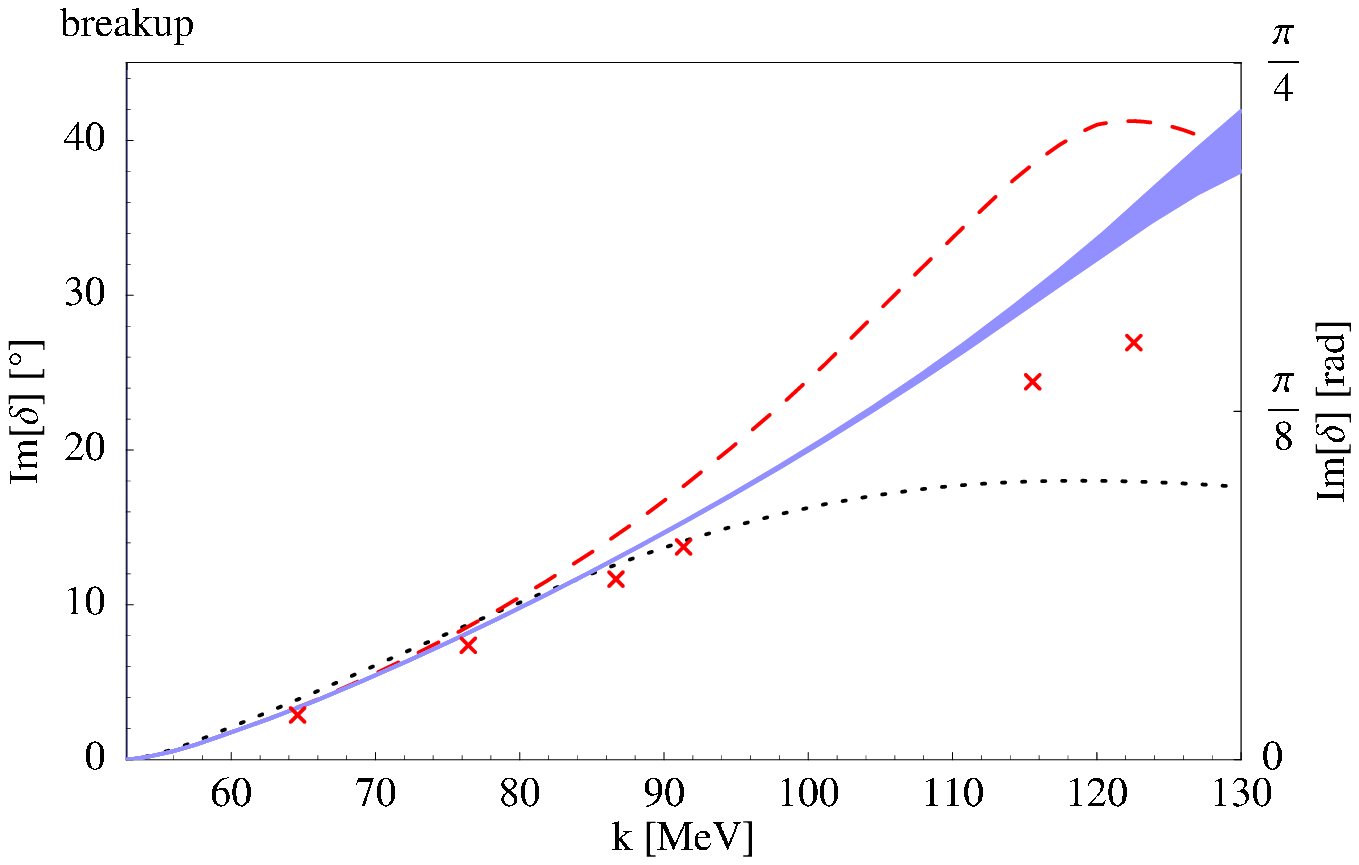}
  \end{center}
  \vspace*{-0pt}
\caption{
  The neutron-deuteron ${}^2\mathrm{S}_\frac{1}{2}$ phase shift at LO (dotted
  line), NLO (dashed line) and NNLO (dark band). Top: Real part.  Bottom:
  Imaginary part.  The band in the NNLO curve corresponds to the change caused
  by a variation of the cutoff from $\Lambda=200$ MeV to $\Lambda=600$ MeV.
  The dots are from the only available direct phase shift analysis of the
  triton channel~\cite{doublet_PSA}, which dates from 1967, and the crosses
  are the results obtained with the Argonne V18$+$Urbana IX two and three-body
  forces~\cite{doublet_pots}.}
\label{doublet_phaseshifts}
\end{figure}
%%%%%%%%%%%%%%%%%%%%%%%%%%%%%%%%%%%%%%%%%%%%%%%%%%%%%%%%%%%%%%%%%%%%

The phase shifts at LO, NLO and NNLO are shown in
Fig.~\ref{doublet_phaseshifts}, together with the result of a phase shift
analysis~\cite{doublet_PSA} and results from numerical calculations using the
two-nucleon Argonne V18 potential and the Urbana IX three-body
force~\cite{doublet_pots}~\footnote{We thank A. Kievsky for providing us with
  his results for the phase shifts at our request.}. The phase shifts converge
steadily in the whole region of validity of our theory, up to centre-of-mass
momenta of the order of $120$ MeV.  At a momentum of the order of $\sqrt{M
  B_3}\sim 90$ MeV ($B_3$ the triton binding energy) the cutoff variation is
of order of $24\%$, $12\%$ and $4\%$ respectively at LO, NLO and NNLO.
Convergence by itself and to the results of the more sophisticated potential
model calculation is achieved, albeit less pronounced in the imaginary part of
the phase shift.

The comparison of our results with those obtained through potential models
must however be done with care.  Results obtained by the use of effective
field theory have an universal validity in the following sense.  Any model
with the same experimental values for the low-energy constants as the ones we
used to fix the low-energy constants of the effective theory, shares also the
same prediction for other observables, \emph{up to the accuracy of the order
  of the EFT calculation}.  In our case, we expect thus that a model like the
Argonne V$18$ two-body force complemented by the Urbana IX three-body force
which predicts or fits the triton binding energy and low-energy two-body phase
shifts, should give phase shifts agreeing with the effective theory result at
the $4\%$, and this is indeed found. The agreement may disappear at higher
order: The effective theory result will depend on low-energy constants
determined by three-body observables that do not enter in the potential model
determination and, through this information, can acquire higher accuracy. This
is especially true when electroweak currents are considered.

A next-order calculation with an accuracy of about $1\%$ is straightforward
and must include the mixing between the ${}^2\mathrm{S}_{\frac{1}{2}}$ and
${}^4\mathrm{D}_{\frac{1}{2}}$ partial waves, the contribution of two-body
$p$-waves and isospin-breaking effects.  However, although it does not involve
any new unknown low-energy constants, it will not provide improved precision,
since of the parameters used to fix our constants at the lower orders, the
scattering length in the doublet $\mathrm{S}$ wave channel is not known with
sufficient accuracy and would dominate the errors.

%%%%%%%%%%%%%%%%%%%%%%%%%%%%%%%%%%%%%%%%%%%%%%%%%%%%%%%%%%%%%%%%%%
\begin{figure}[!htb]
  \begin{center}
    \includegraphics*[width=0.9\linewidth,clip=true]{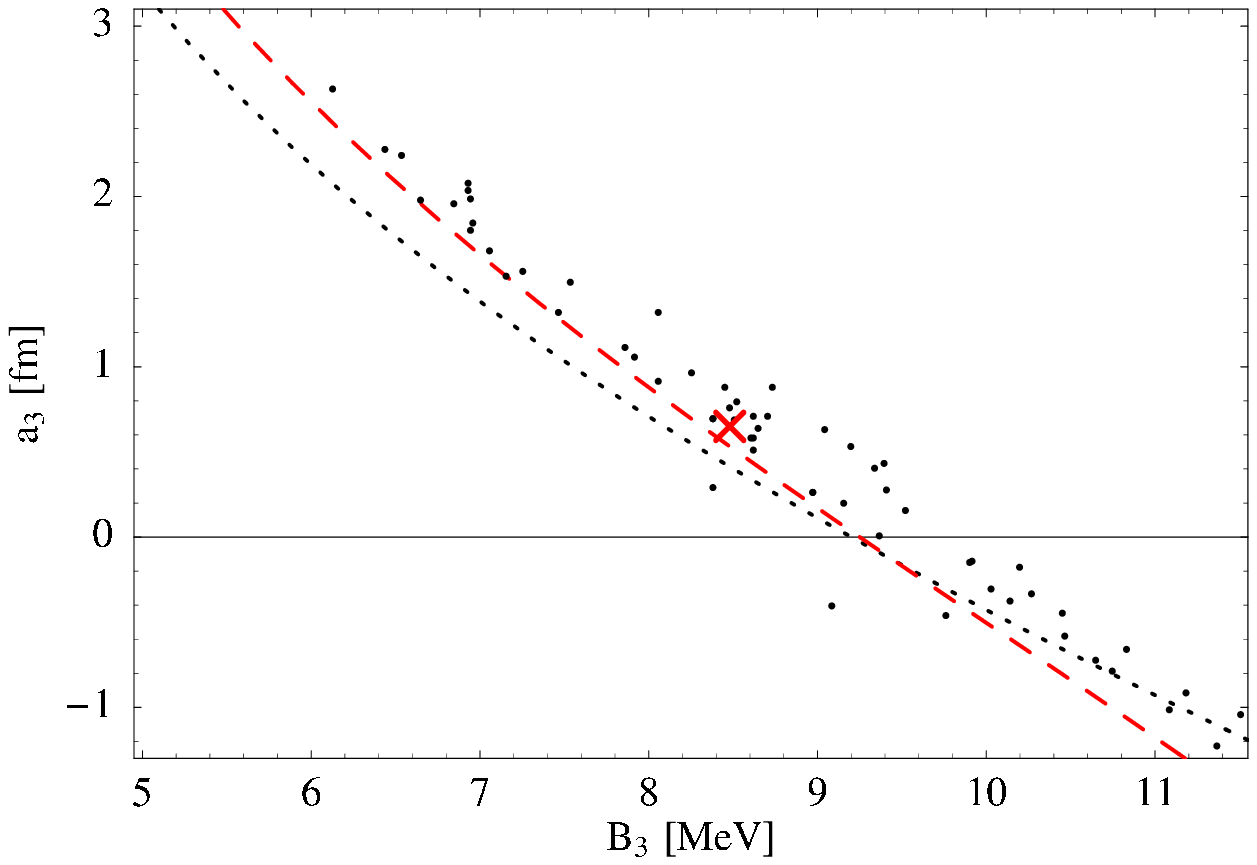}
  \end{center}
  \vspace*{-0pt}
\caption{The Phillips line predicted at NLO. The dots correspond to the
  predictions for the triton biding energy and doublet scattering length in
  different models with the same two-body scattering lengths and effective
  ranges as inputs~\cite{tkachenko}.  The dotted and dashed lines are the EFT
  results at LO and NLO, respectively. The cross is the experimental result. }
\label{phillips}
\end{figure}
%%%%%%%%%%%%%%%%%%%%%%%%%%%%%%%%%%%%%%%%%%%%%%%%%%%%%%%%%%%%%%%%%%%%
The universality of the EFT predictions mentioned above is well illustrated by
the Phillips line~\cite{phillips}.  It is a well-known empirical fact that
different models having the same low-energy two-body physics may have widely
different three-body physics. The variation on the three-body physics is,
however, correlated: One single three-body observable, say the triton binding
energy, determines all the others, up to small corrections. In the EFT
language, this is a reflection of the existence of the parameter $D_0$ (or
$H_0$) already at leading order which is not determined by two-body physics
but whose precise value at fixed cutoff must be determined from another
three-body datum.  Neither $D_0$ nor $\Lambda$ has an immediate physical
significance, but once it is determined at a given cutoff, its cutoff
dependence is fixed by (\ref{master_renorm}). Therefore, as either of these
parameters is varied as the other is kept fixed, the results of different
potential models are reproduced. The basic physics leading to this effect was
explained by Efimov in~\cite{efimovI}. We here exemplify it by showing a plot
of the predictions for the triton binding energy versus the doublet scattering
length in a variety of models in Fig.~\ref{phillips}, together with the EFT
result at LO and NLO.  The NLO result includes the (convenient but
unnecessary) re-summation of higher order terms discussed above. We note that
our NLO result is very similar to Efimov's that was obtained by performing
perturbation theory on the effective range~\cite{tkachenko}.

%%%%%%%%%%%%%%%%%
\section{Conclusions and Outlook}

We presented a simple scheme to classify at which order a given three-body
force must be considered in an EFT approach to the few-body system at very low
energies. These analytical considerations rest on the premise that a three-body
force is included if and only if it is necessary to cancel cutoff dependences
in the observables at a given order.
  
We compared the predictions of this scheme with its numerical implementation
in the case of a NNLO calculation of the $nd$ scattering phase shifts in the
${}^2\mathrm{S}_\frac{1}{2}$ (triton) channel. We confirmed the well-known
result that to LO and NLO, only one momentum dependent three-body force is
necessary for cutoff independence, whose strength can be determined by the
three-body scattering length. We found that one and only one new parameter
enters at NNLO, namely the Wigner-$SU(4)$-symmetric three-body force with two
derivatives. The one additional datum needed to fix it is the binding energy
of the triton.  Indeed, only the $SU(4)$-symmetric three-body forces are
systematically enhanced over their contributions found from a na{\ia}ve
dimensional estimate, with a $2n$-derivative three-body force entering at the
$2n$th order. On the way, we also propose a new, computationally convenient
and simple scheme to perform higher order calculations in the three-body
system by iterating a kernel to all orders, which has been expanded to the
desired order of accuracy in the power counting.
  
The phase shifts thus calculated are in good agreement with a partial wave
analysis and more sophisticated model computations. The result shows
convergence.
  
Work under way includes the NNNLO calculation of the triton channel, its wave
functions, the extension to ${}^3$He and the $pd$ channel, and the inclusion
of electro-weak interactions. This will allow comparison to a plethora of data
and lead to predictions in the range and to an accuracy relevant for
astro-physical problems. Our scheme is clearly extendible to any EFT in which
at least a partial re-summation of graphs at LO is necessary. More formal
aspects of our method must also be addressed.

%%%%%%%%%%%%%%%%%
\begin{acknowledgments}
  This work was supported in part by the Director, Office of Energy Research,
  Office of High Energy and Nuclear Physics, by the Office of Basic Energy
  Sciences, Division of Nuclear Sciences, of the U.S. Department of Energy
  under Contract No.~DE-AC03-76SF00098 (P.F.B.\ and G.R.), by the DFG
  Sachbeihilfe GR 1887/2-1 and the Bundesministerium f{\"u}r Bildung und
  Forschung (H.W.G.), and by the U.S. National Science Foundation under Grant
  No.\ PHY-0098645 (H.-W.H.). H.W.G.~is also indebted to the hospitality of
  the ECT* (Trento) and of the Lawrence Berkeley Laboratory.  We thank A.\ 
  Kievsky for calculating the neutron-deuteron phase shifts above deuteron
  breakup at our request.
\end{acknowledgments}

%%%%%%%%%%%%%%%%%%%%%%%%%%%%%%%%%%%%%%%%%%%%%%%%%%%%%%%%%%%%%%%%%%%%
\begin{appendix}
\section{Asymptotic Amplitude}
In this appendix, we derive the asymptotic form of the half off-shell
amplitudes quoted in the main text.  Consider first the bosonic integral
equation in (\ref{integral_eq_boson}) in the r\'egime $\gamma,\sqrt{ME}\ll p
\ll \Lambda$. For these values of $p$, the integral is dominated by values of
$q$ of the order of $p$. We can then disregard the infrared scales $\gamma$
and $\sqrt{ME}$ as well as the term $H_0/\Lambda^2$ (assuming $H_0\approx 1$).
The integral can be extended to infinity by adding a sub-leading piece.  First
dropping the inhomogeneous term, we find
\begin{equation}\label{asymp}
t(p)=\frac{4}{\sqrt{3}\pi}\frac{1}{p} \int\limits_0^\infty
\dd q\;\;  
\ln\left(\frac{p^2+pq+q^2}{p^2-pq+q^2} \right)
t(q)\;\;.
\end{equation}
A Mellin transformation determines the solution to be a power law $t(p)\sim
p^{s-1}$, where $s$ is determined by the equation
\begin{equation}\label{s}
  I(s)=1
\end{equation}
with
\begin{equation}
I(s)=\frac{4}{\sqrt{3}\pi}\int\limits_0^\infty \dd x\;
\ln\left[\frac{x^2+x+1}{x^2-x+1}\right]
\;x^{s-1}=\frac{8}{\sqrt{3} s}\frac{\sin(\frac{\pi
s}{6})}{\cos(\frac{\pi s}{2})}\;\;.
\end{equation}
The value of $s$ satisfying (\ref{s}) with the largest real part dominates the
asymptotics of $t(p)$.  These values are the imaginary roots $s=\pm \ii
s_0\approx \pm\ii\; 1.0062\dots$.  Because the inhomogeneous term in
(\ref{integral_eq_boson}) behaves like $\sim 1/p^2$ in the asymptotic r\'egime,
it is consistent to ignore it compared to the asymptotics $p^{\pm \ii s_0-1}$
of the homogeneous term. The overall normalisation of $t(p)$ can however not
be determined by this analysis. What is perhaps more surprising is that the
phase of the solution is not determined, either, since both $p^{ \ii s_0-1}$
and $p^{- \ii s_0-1}$ are equally acceptable solutions. Sub-leading
corrections to the asymptotics can be obtained in a similar fashion. For the
corrections suppressed by one power of either $\gamma/p$ or $p\rho$ for
instance, we insert $t(p)=p^{\ii s_0-1} + 8\gamma B_{-1}p^{\ii s_0-2}+B_1 \rho
p^{\ii s_0}$ in the equation
\begin{equation}\label{asymp2}
t(p)=\frac{2}{\pi} \int\limits_0^\infty \dd q\;\frac{q}{p}\;\left(
\frac{2}{\sqrt{3}}\frac{1}{q}         
+\frac{2}{3}\frac{\gamma}{q^2}       
+\frac{r}{2}\right)  
\ln\left(\frac{p^2+pq+q^2}{p^2-pq+q^2} \right) t(q)
\end{equation}
and keep terms linear in $\gamma$ and $r$. Matching order by order, we find
\begin{equation}
B_n=\frac{\sqrt{3}}{4}\frac{I(\ii s_0+n)}{1-I(\ii s_0+n)}\;\;.
\end{equation}
Taking real combinations of the solutions corresponding to $s_0$ and $-s_0$,
we finally arrive at the result quoted in (\ref{NLO_approx_t}). Proceeding in
a similar way, one can obtain this kind of correction to all orders, if
necessary. We were however unable to find the sum of this series.
\end{appendix}

%%%%%%%%%%%%%%%%%%%%%%%%%%%%%%%%%%%%%%%%%%%%%
\end{section}

%%%%%%%%%%%%%%%%%%%%%%%%%%%%%%%%%%%%%%%%%%%%%%%%%%%%%%%%%%%%%
\bibliographystyle{apsrev}

\end{document}